\documentclass[12pt,a4paper,twoside]{article}
\usepackage{epsfig}
\newlength{\figwidth}
\newlength{\figheight}
\newlength{\doublefigheight}
\newlength{\twofigheight}
\def\gaga{$\gamma \gamma$}

\def\wgg{$W_{\gamma \gamma}$}

\def\ggg{$\Gamma_{\gamma \gamma}$}

\def\pgg{$\phi_{\gamma \gamma}$}

\def\epem{$e^+ e^-$}

\def\z0{Z}

\def\tgb{$\tan \beta$}

\title{ \vspace*{2cm}
      Determination of the Higgs-boson  couplings and $H-A$ mixing
         in the generalized SM-like Two Higgs Doublet Model
}

\author{P.Nie\.zurawski, A.F.\.Zarnecki \\
{\small\it Institute of Experimental Physics, Warsaw University,} \\
        M.Krawczyk \\
{\small\it Institute of Theoretical Physics, Warsaw University} 
} 

\begin{document}

\maketitle

\vspace*{2cm}

\begin{abstract}

The feasibility of measuring the Higgs-boson properties at 
the Photon Collider at TESLA has been studied in detail for masses 
between 200 and 350~GeV,
using realistic luminosity spectra and detector simulation. 
We consider  the Two Higgs Doublet Model~(II)
with SM-like Yukawa couplings for $h$,
 parametrized by only one parameter (\tgb).
The combined measurement of the invariant-mass 
distributions in the $ZZ$ and $W^+ W^-$ decay-channels is
sensitive to both the two-photon width \ggg\ and phase \pgg.
From the analysis  including systematic uncertainties
we found out that after one year of Photon Collider running 
with nominal luminosity
the expected precision in the measurement of \tgb\
is of the order of 10\%, for both light ($h$) 
and heavy ($H$) scalar Higgs bosons.
The $H-A$ mixing angle $\Phi_{HA}$, 
characterizing a weak CP violation in the model with two Higgs doublets, 
can be determined 
to about 100~mrad, for low $\tan \beta$.
\end{abstract}

\thispagestyle{empty}

\clearpage

%
%

\section{Introduction}
\label{sec:intro}

A photon collider has been  proposed as a natural
extension of the \epem\ linear collider \cite{pc}.
The physics potential of a photon collider is very
rich and complementary to the physics program of the \epem\
and hadron-hadron colliders.
It is an ideal place to study the mechanism of the electroweak 
symmetry breaking (EWSB) and the properties of the Higgs-boson.
In paper~\cite{nzk_wwzz} we have performed realistic simulation of 
the production  at the TESLA Photon Collider~\cite{tdr_pc}
of the SM Higgs boson with masses above 150~GeV
for $W^+ W^-$ and $\z0 \z0$ decay channels.
We found that precise measurements of both 
the $higgs \rightarrow \gamma \gamma$ partial width, 
\ggg\ and the phase, \pgg\ are needed for determination 
of the Higgs-boson couplings,
see also \cite{ginzburg,cros_zz,gounaris,asakawa,hagiwara}.
Therefore, as we have found in \cite{nzk_wwzz}, it is extremely important
to combine both  $W^+ W^-$ and $ZZ$ channels, as the first one, 
due to a large background leading to large interference effects 
is very sensitive to a phase, while the second one - 
to a partial width.
From combined analysis of  $W^+ W^-$ and $\z0 \z0$ decay channels
the $\gamma \gamma$ partial width \ggg\ can be measured 
with an accuracy of 3 to 8\%
and the  phase of the $\gamma \gamma h$ amplitude 
\pgg\ with an accuracy  between 30 and 100~mrad.

Once we found how useful these channels are for the SM Higgs study,
the natural is to investigate models with the extended Higgs sector.
In this paper we continue the analysis from \cite{nzk_wwzz} 
by extending it to the Two Higgs Doublet Model 2HDM~(II).
We consider one of the SM-like versions of this model,
with Yukawa couplings as in SM (up to a sign).
For  both CP-conserving as well as CP-violating scenarios
we perform the combined analysis of $W^+ W^-$ and $ZZ$ invariant-mass 
distributions and extract the corresponding Higgs-boson coupling
to gauge bosons, which is governed by only one parameter ($\tan\beta$).
For 2HDM model with a CP violation we estimate the precision 
of the measurement of $H-A$ mixing angle. 
The systematic uncertainties of the coupling and mixing
angle measurements are estimated.
Results given in this paper supersede preliminary results 
presented in the first part of our earlier paper \cite{eps2003}.

%
%

\section{Event simulation}

The Compton back-scattering of a laser light off  high-energy
electrons is considered as a  source of high energy, highly polarized
photon beams at Photon Collider~\cite{pc}. 
According to the  current design \cite{tdr_pc}, the energy
of the laser photons is assumed to be fixed
for all considered electron-beam energies.
With 100\% circular polarization of laser photons 
and 85\% longitudinal polarization of the electron beam
the luminosity spectra peaked at high $\gamma \gamma$ invariant masses
is expected.

Our analysis uses the CompAZ parametrization \cite{compaz} of
the realistic luminosity spectra for a Photon Collider 
at  TESLA \cite{telnov}.
We assume that the centre-of-mass energy of colliding electron beams, 
$\sqrt{s_{ee}}$, is optimized for the production of
a Higgs boson with a given mass.
All results presented in this paper were obtained 
for an integrated luminosity corresponding to one year 
of the photon collider running, as given by \cite{telnov}.
The total photon-photon luminosity increases from about
600~fb$^{-1}$ for $\sqrt{s_{ee}}=305$~GeV 
(optimal beam energy choice for $M=200$~GeV) to
about 1000~fb$^{-1}$ for $\sqrt{s_{ee}}=500$~GeV
($M=350$~GeV).

The analysis described in this work was performed in two steps.
In the first step, we use samples of events generated with
PYTHIA~6.152 \cite{PYTHIA} 
to study selection efficiency and invariant-mass resolution 
in reconstruction of 
the events
$\gamma \gamma \rightarrow W^+ W^- / \z0\z0 $,
as a function of the \gaga\ centre-of-mass energy, \wgg .
We consider the  direct vector-boson production 
in \gaga\ interactions as well as Standard Model Higgs-boson decays 
into the vector bosons and contribution from the interference terms.
To take into account effects which are not implemented in PYTHIA
(photon beam polarization, interference term contribution,
direct $\gamma \gamma \rightarrow \z0  \z0 $ production)
we use the standard method used in various experimental analyses
 called a reweighting procedure.
Each generated event is attributed a weight given 
by the ratio of the differential cross-section for a vector-boson 
production in the polarized photon interactions \cite{ginzburg,cros_zz,cros_ww}
to the PYTHIA differential cross section for given event.
The fast simulation program SIMDET version 3.01 \cite{SIMDET}
is used to model the TESLA detector performance.

A good invariant-mass resolution is essential for the
considered measurement. 
For the $W^+ W^-$ events only $W^+ W^- \rightarrow qq\bar{q}\bar{q}$ 
decay channel is considered, 
as without knowing the exact beam-photon energies,
which is always a  case for the Photon Collider,
the semileptonic $W^\pm$ decays worsen the mass resolution.
The final selection efficiency for 
$\gamma \gamma \rightarrow W^+ W^-$ events 
is found to lay between 20\% for \wgg \,$\sim$ ~200 GeV  and 16\% 
for \wgg \,$\sim$ ~400 GeV (including 47\% probability for 
hadronic decays of both $W^\pm$).
The resolution in the reconstructed $\gamma \gamma$
invariant mass for these events,
described by the parameter $\Gamma$ (from 
the Breit--Wigner type fit), changes from
about 6.5 GeV at \wgg \, = ~200~GeV to about 13 GeV at \wgg \,= ~400~GeV.

For  $\z0 \z0$ events, only $\z0 \z0 \rightarrow l\bar{l}q\bar{q}$ decay
channel is considered, where one $\z0$ decays into $e^+ e^-$ or 
$\mu^+ \mu^-$.
Lepton tagging and the invariant-mass reconstruction for
both the lepton pair and the two hadronic jets is crucial
for a suppression of the background from the direct
$\gamma \gamma \rightarrow W^+ W^-$ events.
After all cuts, the selection efficiency for $\z0  \z0 $ events 
is only about 5\%,
mainly due to a small branching ratio for the considered channel
(9.4\% for  $\z0 \z0 \rightarrow l\bar{l}q\bar{q}, \; l=e,\mu$),
however, the final sample is very clean.
For the $l\bar{l} q \bar{q}$ final state we get the invariant-mass 
resolution $\Gamma$ changing from
about 5.5 GeV at \wgg \, = ~200 GeV to about 7.5 GeV at \wgg \, = ~400 GeV.
Details of event selection and the description of the
invariant mass resolution are given in \cite{nzk_wwzz}.

The invariant-mass resolutions obtained from a full simulation of $W^+ W^-$  
and $\z0 \z0$ events (based on the PYTHIA and SIMDET programs), 
have been parametrized as a function of the \gaga\ centre-of-mass energy, \wgg .
They can be used to obtain the parametric description of 
the expected invariant mass distributions, 
for $\gamma \gamma \rightarrow W^+ W^-$ 
and $\gamma \gamma \rightarrow \z0  \z0 $ events,
avoiding time consuming event generation procedure.
For arbitrary model, and for arbitrary values of model parameters, 
expected mass distributions
can be calculated by numerical convolution of the parametrized 
resolutions with 
the CompAZ spectra and the cross section formula. 
This approach, developed in  \cite{nzk_wwzz}, was used to obtain
results described in the remaining part of this paper.

%
%

\section{Measurement of \boldmath{\ggg} and \boldmath{\pgg} 
         from the $WW/ZZ$ \\ invariant-mass distributions in SM} 
\label{sec:width}

In this section we summarize results of \cite{nzk_wwzz},
where the feasibility of measuring the Standard Model Higgs-boson 
$W^+ W^-$ and $\z0 \z0$ decay channels  at the \gaga\ option of TESLA
has been studied for a Higgs-boson mass above 150~GeV.
The signal, i.e. the Higgs-boson decays into the 
vector bosons, and the background from direct vector-bosons production
are used to extract the width and the phase of the loop
coupling $ higgs \rightarrow \gamma \gamma$.
For the $\z0  \z0 $ final-state a direct, i.e. non-resonant, 
process is rare as it occurs via  loop only.  
On contrary, the non-resonant $W^+W^-$ 
production is a tree-level process, and is expected to be large.
Therefore, also an interference between the signal of $W^+W^-$ production 
via the Higgs resonance and the background from the direct  $W^+W^-$ production
may be large.
This effect can be used 
to access an information about the phase  \pgg.
For the Higgs-boson masses around 350~GeV we found in \cite{nzk_wwzz}
that the phase \pgg\ is more sensitive 
to the loop contributions of new,
heavy charged particles than  the width \ggg\ itself.

The parametric description of the expected invariant mass distributions 
for $\gamma \gamma \rightarrow W^+ W^-$ 
 and $\gamma \gamma \rightarrow \z0  \z0 $ events
was  obtained by the numerical convolution 
of the cross-sections formula (including \ggg\ and \pgg\ as model parameters)
with the CompAZ spectra and the parametrized resolution.
Based on this description,
many  experiments were simulated, each corresponding to one year
of a Photon Collider running at TESLA at a nominal luminosity.
The ``theoretical'' distributions were then fitted, simultaneously to
the ``observed'' $W^+ W^-$ and $\z0 \z0$ mass spectra, with the 
width \ggg\ and phase \pgg\ considered as the only free parameters.
Assuming the Standard Model Higgs-boson branching ratios, and
with a proper choice of the electron-beam energy, the $\gamma \gamma$
partial width can be measured with an accuracy of 3 to 8\%, while
the  phase of the amplitude with 
an accuracy between 35 and 100~mrad~\cite{nzk_wwzz}.

The \pgg\ measurement opens a new window to a precise
determination of the Higgs-boson couplings
and to  search for a ``new physics''.
It turns out that the phase is constrained predominantly by the $W^+ W^-$
invariant-mass distribution, thanks to the large interference between
Higgs-boson decay into  $W^+ W^-$  and non-resonant (SM) $W^+ W^-$ production.
However, the two-photon width of the Higgs-boson is much better constrained
by the measurement of the $\z0 \z0$ mass spectra, as the non-resonant
background is much smaller here.
A precise determination of both parameters is only possible
when both measurements are combined.

The promising results obtained for the SM Higgs boson
encourage to the evolution of the analysis towards
the models with more Higgs doublets.
Presented analysis extends the approach 
developed for SM Higgs boson to the selected 
Two Higgs Doublet Model~(II) scenarios. 
For all considered sets of parameter values
the expected invariant mass distributions were calculated,
taking into account model predictions for both the production 
cross-section (including interference term contributions) 
and  the Higgs-boson branching ratios.
Parametrized distributions were used to simulate many experiments,
each corresponding to one year of a Photon Collider running.
Errors (and correlations) in \ggg\ and \pgg\ measurement,
expected from the simultaneous fit 
to the observed $W^+ W^-$ and $\z0 \z0$ mass spectra,
where then used to determine expected 
uncertainties of model parameters.

%
%

\section{Determination of the Higgs-boson couplings  \\
       in the CP-conserving model $B_h$}
\label{sec:2hdm}

In our previous analysis \cite{nzk_wwzz}
we indicated possible deviations from the Standard Model predictions
resulting from the loop contributions to the $h \gamma \gamma$ vertex
 of new heavy charged particles. 
However, deviations in the two-photon width and phase
can also appear if the couplings of the Higgs-boson to the other particles are 
different than those predicted by the Standard Model.
Both such effects are taken into account in this analysis,
based on a simple extension of SM, namely 2HDM~(II), see also \cite{2HDM}.

In this section we consider the Two Higgs Doublet Model~(II) (2HDM~(II)) 
with CP conservation.
The Higgs sector of such model contains $h$, $H$, $A$ and $H^\pm$ bosons,
and is characterized (in the $Z_2$ symmetric case)
by mixing angles $\alpha$ and $\beta$.

%
%

\subsection{The Model and its tests at future colliders}

The simplest version of the 2HDM is the case where couplings are
parametrized by one mixing angle.
Often one considers 2HDM when $h$ couples to gauge bosons as in SM - 
however then also all Yukawa couplings are as in SM.
Here we assume instead that the Yukawa couplings of $h$ are equal (up to sign) 
to the corresponding SM Higgs-boson couplings. 
Then the coupling of $h$ to gauge bosons is governed by \tgb, 
as it is  shown in table~\ref{tab:2hdm}.
The corresponding Yukawa and gauge boson couplings 
of $H$ and $A$ bosons  are also
uniquely determined by \tgb, as can be seen in the table.
Note that here $H$ and $A$ have similar Yukawa couplings
(up to $i \gamma_5$ factor).\footnote{%
In our analysis the loop  $h \gamma \gamma$ and  $H \gamma \gamma$  
vertexes appear. The couplings of $h$ and $H$  to the charged higgs 
boson $H^\pm$, contributing to such vertexes, are calculated 
according to the 2HDM II potential \cite{2HDM} assuming $\mu$=0.
}
This scenario, called by us a  model $B_h$, 
can be treated as a generalized Standard Model-like 2HDM~(II) 
scenario $B$ for $h$, introduced in \cite{2HDM}.
Solutions $B_{h+u}$ and $B_{h-d}$ of \cite{2HDM}
correspond to our model $B_h$ in the limit of $\tan \beta \ll 1$ 
and $\tan \beta \gg 1$, respectively.

In 2HDM there is a possibility that one of the Higgs boson couplings 
vanishes, for example the coupling to the EW gauge bosons. 
The corresponding sum rule
\begin{eqnarray}
(\chi^h_X)^2+( \chi^H_X)^2+(\chi^A_X)^2 & = & 1, \nonumber
\end{eqnarray}
where $X$ denotes a fermion or a vector boson, $X=u,\; d, \; V$,
ensures only that at least one of the other Higgs 
bosons has a nonzero coupling.
In the solution $B_h$ 
the lightest Higgs-boson $h$ couplings to fermions 
are equal to the Standard Model,
except that for up-type fermions the sign is opposite.\footnote{
This is equivalent to the following condition
imposed on the mixing angles $\alpha$ and $\beta$ of 2HDM~(II):
$ \alpha = -\frac{\pi}{2} - \beta $.
} 
In such a case, the two-gluon partial width, $\Gamma_{gg}$,
dominated by the top-quark loop contribution, is very close
to the SM predictions.
\renewcommand{\arraystretch}{1.3}
\begin{table}[b]
\begin{center}
\begin{tabular}{|l|c|c|c|}
\hline
    & $h$ & $H$ & $A$ \\ \hline
$\chi_u $  & $-1$ & $-\frac{1}{\tan\beta}$ &  
  $-i  \; \gamma_5  \; \frac{1}{\tan\beta} $ \\
$\chi_d $  & $+1$ & $-\tan\beta$ &  
  $-i  \; \gamma_5  \; \tan\beta $ \\
$\chi_V $  & $\cos ( 2 \beta)$ & $- \sin ( 2 \beta)$ &    $0$ \\
\hline
\end{tabular}
\end{center}
\caption{
Couplings of the neutral Higgs-bosons to up- and down-type fermions,
and to vector bosons,
relative to the Standard Model couplings, for the considered solution $B_h$ 
of the SM-like 2HDM~(II).
}
\label{tab:2hdm}
\end{table}

We observe that couplings of $h$ to the EW gauge-bosons $V$, 
$V=W^\pm, Z^\circ$,
may significantly differ from the corresponding SM couplings. 
However, since the decays into $WW$ and $ZZ$ are the dominating ones
in the considered Higgs boson mass range\footnote{
For the SM Higgs boson with mass of 200 to 350 GeV 
contribution of other decay channels is about 0.4\%.} 
this potentially large 
effect cancels out in the branching ratios  
(as it affects in similar strength both the partial width 
and the total width). 
Therefore, to be sensitive to deviations
from the SM couplings, one has to study processes where the 
coupling $hVV$ is involved in the production of the Higgs boson
and not only in its decays. 
Taking this into account
we expect that, i.e. $gg \rightarrow h \rightarrow VV$
at LHC could indicate no deviations from SM,\footnote{
In solution $B_h$, deviations of more than 10\% 
from SM predictions (on the number of $WW$ and $ZZ$ decays) 
are only expected for $|\chi_V| < 0.2$, 
i.e. for  $\tan \beta \approx 0.8 - 1.2$. 
} 
and at the same time sizable effect can appear at LC 
(eg. in Higgsstrahlung process $e^+e^- \rightarrow  Z h$).
Also one can expect large deviations from the SM prediction 
in the two photon width $\Gamma_{\gamma \gamma}$ and phase 
$\phi_{\gamma \gamma}$ which can be 
measured  at the Photon Collider, as described below.

Finally we note that when the Higgs boson coupling to V vanishes, 
so does the corresponding branching ratio
and the  sensitivity of the considered measurements is lost. 
This is a reason why the determination of $\tan \beta$
from measurement of $h$ production and decays to  $W^+ W^-$ and $\z0 \z0$
is not possible for $\tan \approx 1$.

%
%

\subsection{  $\gamma \gamma \rightarrow h/H$   in the model $B_h$ }

In the Standard Model, the dominant contributions to
the $h \gamma \gamma$ coupling are due to the
$W^\pm$ and top-quark loops.
Therefore, the process of resonant Higgs boson production
at the Photon Collider is  sensitive to the 
Higgs-boson couplings to the gauge-bosons.
Moreover, as the phase of the $W^\pm$ contributions to the 
$\gamma \gamma \rightarrow h $ amplitude differs from that
of the top loop contribution, not only the two-photon
partial width \ggg\ is sensitive  to the the Higgs boson 
coupling, but also the phase \pgg.
Both parameters can be precisely measured at the Photon 
Collider (see Section \ref{sec:width}), allowing
us to constrain the values of $\chi_V$ and $\tan \beta$ from the  
combined measurement of $W^+ W^-$ and $\z0 \z0$ invariant-mass
distributions.
An important observation is that the ratio
of $h$ and $H$ branching ratios to the vector bosons,
BR$(higgs \rightarrow ZZ)$/BR$(higgs \rightarrow W^+ W^-)$ 
does not depend on the $\tan\beta$ value and for a given Higgs-boson mass 
it is expected to be the same as in the Standard Model.

For $\tan \beta \ll 1$ one gets $\chi_V^h \approx 1$. 
Nevertheless, 
one expects significant deviations from the Standard Model predictions 
for a light Higgs-boson $h$, both for the two-photon width and phase,
since as compared to the SM there is a change of a relative sign of the 
top-quark and the $W$ contributions.
In 2HDM II (sol. $B_h$) the two-photon width  is significantly larger than
in the Standard Model, where these two contributions partly cancel each other.
For $\tan \beta \sim 1$ the two-photon width decreases, due to the
suppressed $W$-loop contribution  (here $\chi_{V} = \cos (2\beta)  \approx 0$).
Finally, for large values of $\tan \beta$ ($\cos( 2 \beta)\approx -1$),  
the two-photon width of the light Higgs-boson $h$ tends to be  close 
to the expectations of the Standard Model, since the only difference
is due to the presence of the heavy charged Higgs-boson in the loop.
The opposite, as compared to the SM one,
sign of the down-type fermion contributions gives negligible effect.

The two-photon width and phase
can be  investigated also for the heavy scalar Higgs-boson $H$
for  model  $B_h$, 
with couplings as given in table \ref{tab:2hdm}.

%
%

\subsection{Determination of \tgb\ }

\begin{figure}[tb]
  \begin{center}
     \epsfig{figure=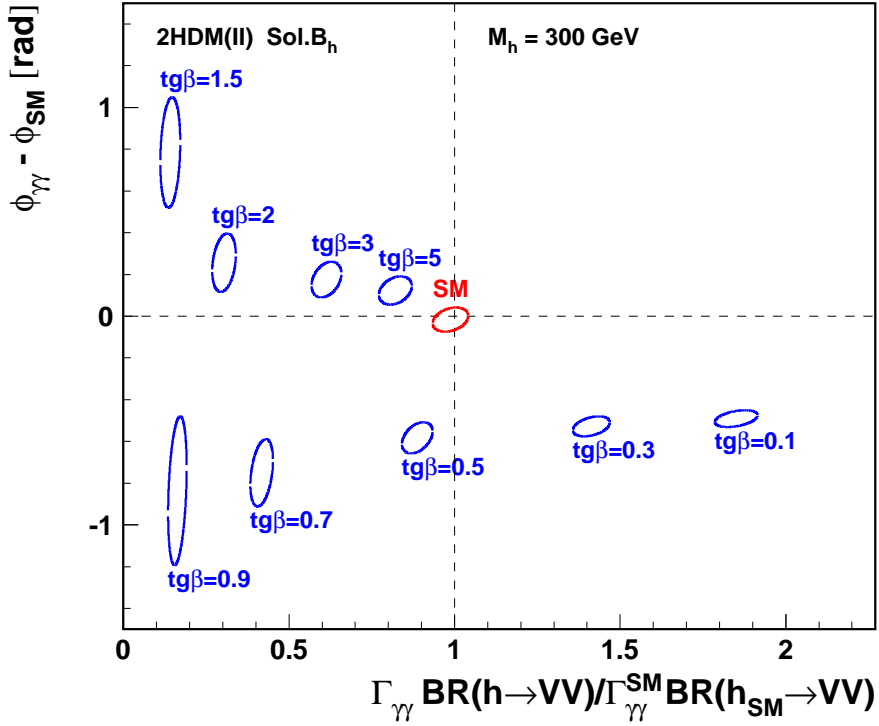,height=\figheight,clip=}
  \end{center}
 \caption{ 
   The deviation from the SM for the light Higgs-boson $h$ with mass 
300~GeV in the SM-like 2HDM~II (sol. $B_h$), 
with charged Higgs-boson mass of 800~GeV for different 
values of $\tan \beta$.
Statistical error contours (1$\sigma$) on the measured deviation from
the Standard Model predictions for
the phase \pgg\ and for
$\Gamma_{\gamma \gamma} \; \times \; BR(h\rightarrow VV)$,
correspond to $L_{\gamma \gamma} \approx 840$~fb$^{-1}$. 
Contour labeled 'SM' indicates the expected precision for the Standard 
Model.
 } 
 \label{fig:showel1} 
 \end{figure}

\paragraph{Results for $h$ production}
Results  for the 
light Higgs-boson $h$ with mass $M_h = 300$ GeV, 
 from the measurement of the two-photon 
width (times the vector-boson branching ratio) and phase 
are presented in Fig.~\ref{fig:showel1} for various $\tan \beta$ values
and the charged Higgs-boson mass of 800~GeV.
Statistical error contours (1$\sigma$) on the expected deviation from
the Standard-Model predictions
are obtained from  the combined fit to the invariant-mass distributions for 
$W^+ W^-$ and $Z Z$ events.
They correspond to  $L_{\gamma \gamma} \approx 840$~fb$^{-1}$. 
These contours show that the measurement
of the two-photon width and phase for the light Higgs-boson $h$
decaying into $W^+ W^-$ and $Z Z$ would allow  a precise determination 
of the $\tan \beta$ value. 
The possible ambiguity in the measurement of the two-photon width,
observed for low value of $\Gamma_{\gamma \gamma} \cdot BR(h \rightarrow VV)$,
can be resolved by the phase measurement, 
which clearly distinguishes between
low $\tan \beta$ and large $\tan \beta$ solutions.

%
%
%

%
\begin{figure}[tb]
  \begin{center}
     \epsfig{figure=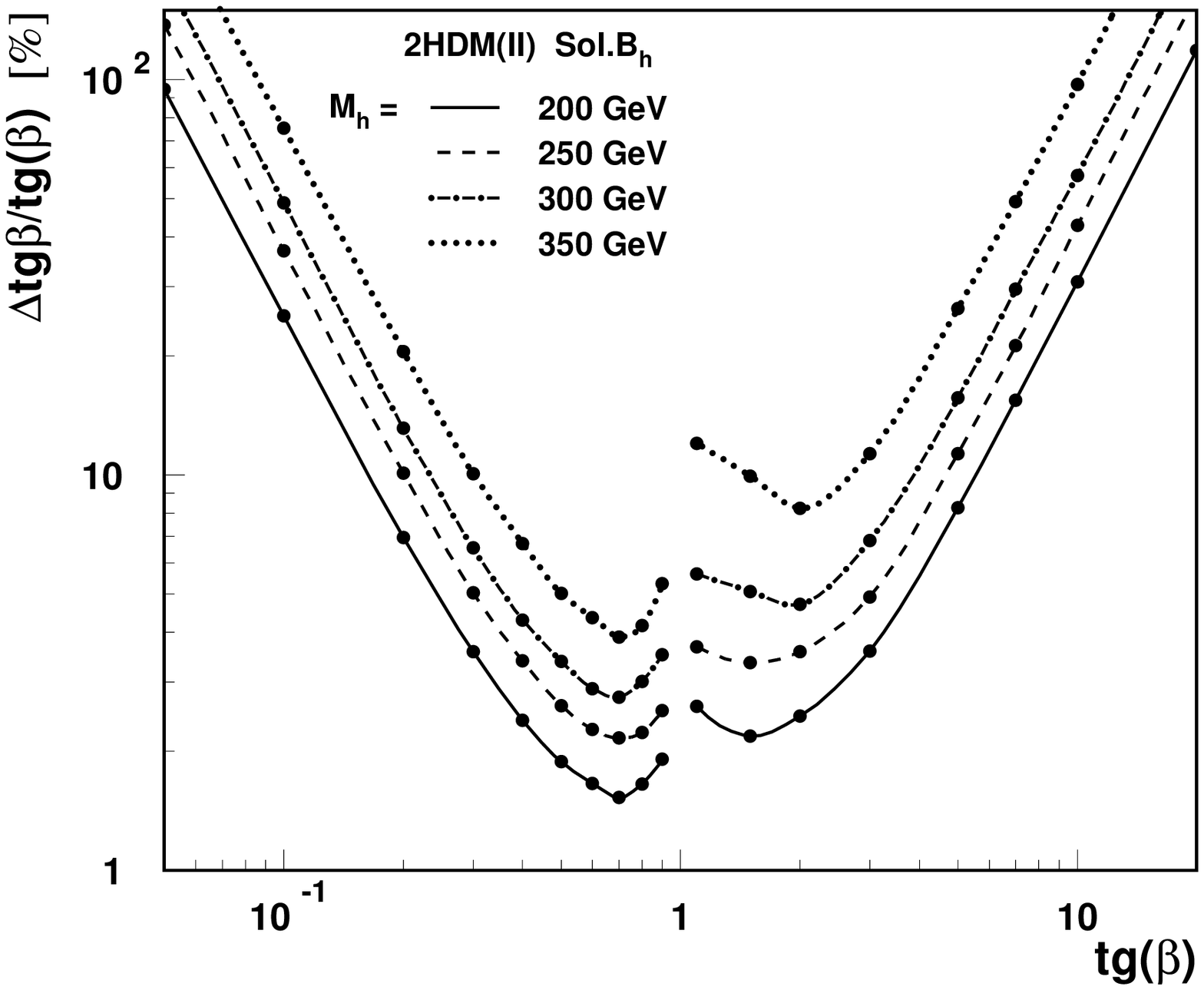,height=\figheight,clip=}
  \end{center}
 \caption{ 
          Statistical error in the determination of  $\tan \beta$,
          for four values of the light Higgs-boson mass $M_h$. 
The simultaneous fit to the observed $W^+ W^-$ and  $ZZ$ mass spectra 
is considered for the 
SM-like 2HDM~II (sol. $B_h$), 
with charged Higgs-boson mass of 800~GeV.
Centre-of-mass energy of colliding electron beams $\sqrt{s_{ee}}$ 
is optimized for each mass $M_h$.
           } 
 \label{fig:errlistat} 
 \end{figure} 
%

%
The statistical error on the extracted  $\tan \beta$ value
is shown in Fig.~\ref{fig:errlistat} for different values 
of the light Higgs-boson mass $M_h$.
The expected error in the $\tan \beta$ determination is smallest 
(from about 1.5\% for $M_h=200$~GeV to about 4\% for $M_h=350$~GeV)
for $\tan \beta \approx 0.7$, i.e. close to 1.
Although the resonant production cross section is small
in this region the Higgs-boson coupling to the vector bosons
is most sensitive to  $\tan \beta$.
For very high and very low $\tan \beta$, when the
relative  Higgs-boson coupling to the vector bosons is close to $\pm1$ 
(table \ref{tab:2hdm}),
the precise measurement of $\tan \beta$ is not possible.
For $\tan \beta \approx 1$ no direct measurement 
and no error estimate is possible,
as the coupling to the vector bosons vanishes. 
In such a case only limits on the $\tan \beta$ value can be set
within the considered model.

\begin{figure}[p]
  \begin{center}
     \epsfig{figure=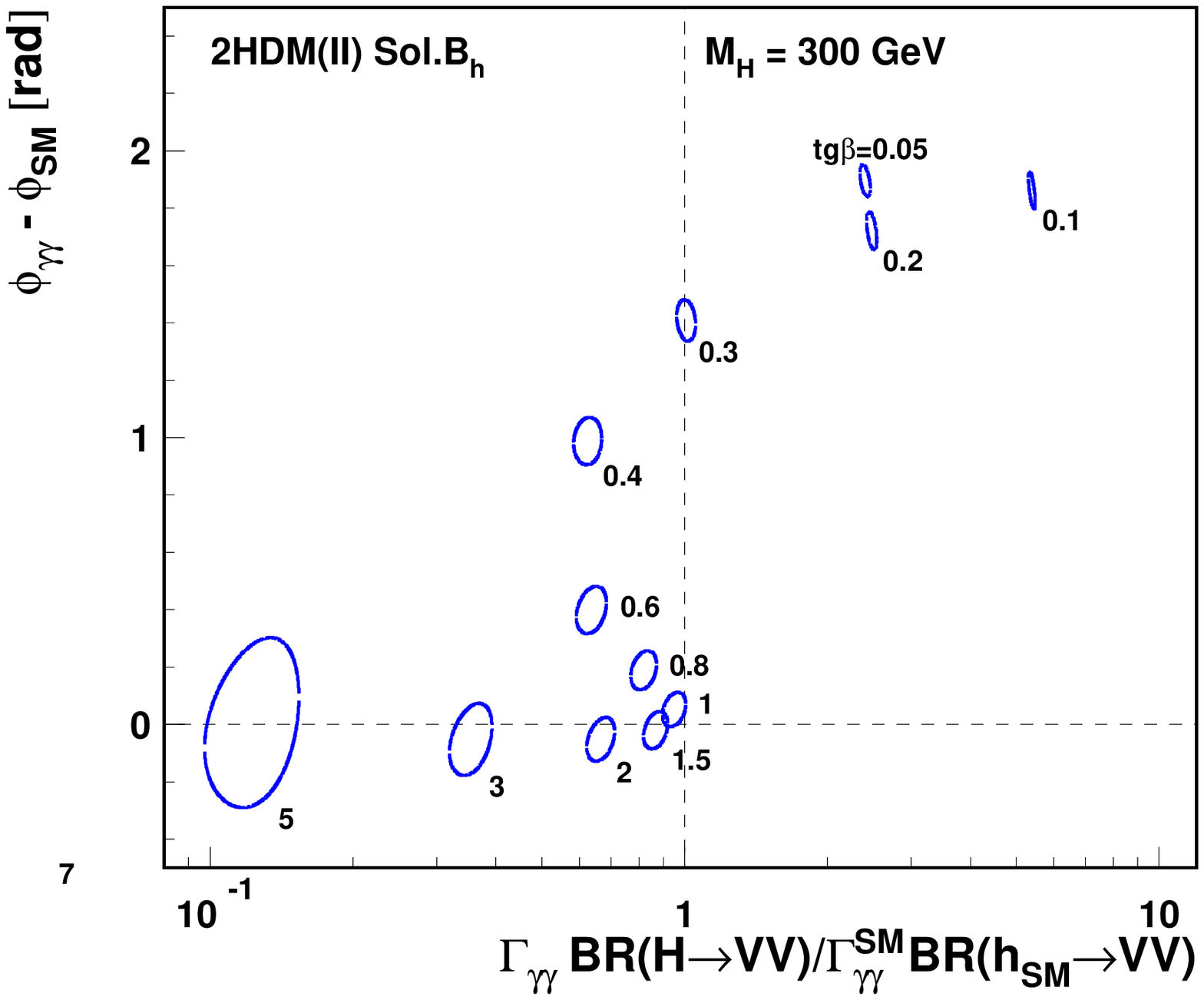,height=\doublefigheight,clip=}
  \end{center}
 \caption{ 
     As in Fig. \ref{fig:showel1} for     the heavy
 Higgs-boson $H$ with mass $ 300$ GeV.
A light Higgs-boson mass is assumed to be $M_h = 120$~GeV.
         } 
 \label{fig:showel2} 
 \end{figure}

\paragraph{Results for $H$ production}
The measurement of the two-photon width and phase
has been  investigated also for the heavy scalar Higgs-boson $H$
of the SM-like Two Higgs Doublet Model~(II)  sol. $B_h$ (as before), 
with couplings as given in table \ref{tab:2hdm}.
Statistical error contours (1$\sigma$) on the expected deviations from
the Standard Model predictions are presented in Fig.~\ref{fig:showel2},
for the heavy scalar  $H$ with mass $M_H = 300$~GeV,
while a light Higgs-boson mass is set to 120~GeV and that of the 
charged Higgs-boson  to 800~GeV.
For $\tan \beta \sim 1$ both the two-photon width and phase of  $H$ 
are  close to the expectations of the Standard Model (for a given $M_H$). 
For  $\tan \beta > 1$ both the top-quark and $W$  
contributions to the two-photon width are strongly suppressed
and the precision of the measurement deteriorates fast
with increasing $\tan \beta$.
For $\tan \beta < 1 $   the $W$-loop contribution
decreases with decreasing $\tan \beta$, however the top-quark 
contribution to the two-photon width increases.
As a result, the two-photon width decreases slightly for 
$\tan \beta \sim 0.5$ and then starts to increase with 
decreasing $\tan \beta$.
For $\tan \beta \sim 0.1$ the  Higgs-boson decay to $c \bar{c}$
starts to dominate. The expected number of events with the $W^+ W^-$ and $ZZ$
decays drops rapidly and the measurement becomes problematic again.

%
%

%
\begin{figure}[p]
  \begin{center}
     \epsfig{figure=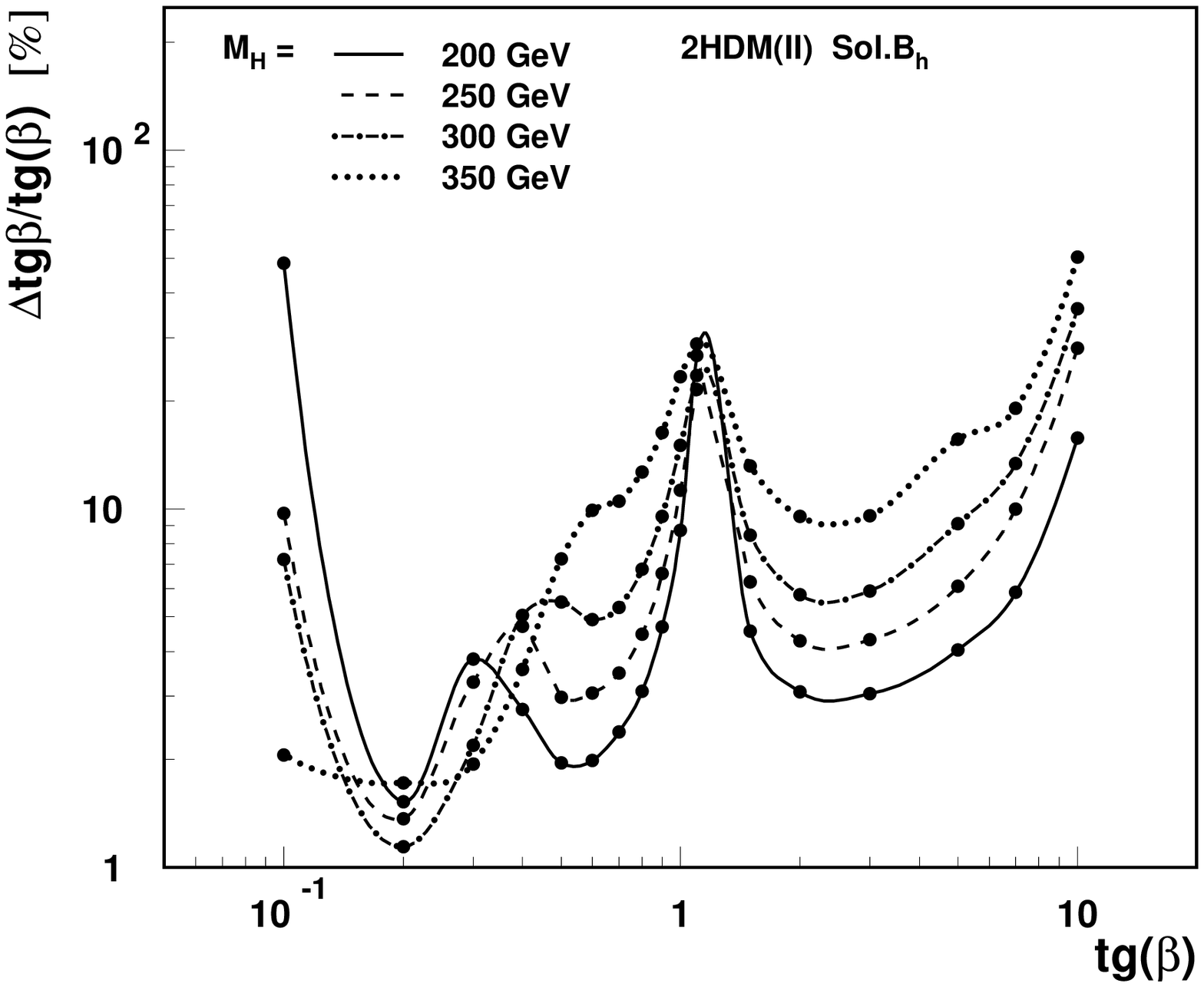,height=\doublefigheight,clip=}
  \end{center}
 \caption{ 
As in Fig. \ref{fig:errlistat}
          for a  heavy Higgs boson $H$, 
with a light Higgs-boson mass of $M_h = 120$~GeV.
} 
 \label{fig:errhvstat} 
 \end{figure} 
%

%
The statistical error on the extracted  $\tan \beta$ value
is shown in Fig.~\ref{fig:errhvstat}.
Results 
are given for four  values of heavy scalar Higgs-boson mass $M_H$,
 from 200 to 350 GeV.
The expected error in the $\tan \beta$ determination is smallest
(1--2 \%) for $\tan \beta \approx $0.2.
For larger values of $\tan \beta$,  
$0.3 \le \tan \beta \le 0.8$, 
the precision depends strongly on the
Higgs-boson mass. 
For mass of 200~GeV it changes between 2 and 4\%,
whereas for mass of 350~GeV it is between 2 and 10\%.
We checked that the precise measurement is also possible 
for   $1.5 \le \tan \beta \le 5$,
with statistical errors from 3--4\% for $M_H=$200~GeV  to
10--20\% for $M_H=$350~GeV.


\section{Determination of $H-A$ mixing 
 for model $B_h$ with a weak CP violation}
\label{sec:cphiggs}

In the  general  Two Higgs Doublet Model \cite{CP2HDM},
the mass eigenstates of the neutral Higgs-bosons
$h_1$, $h_2$ and $h_3$
do not  match CP eigenstates $h$, $H$ and $A$.
We consider here the CP-violating 
Two Higgs Doublet Model based on solution $B_h$,
with a weak CP violation
through a small mixing between $H$ and $A$ states.
We study a simple option, where
the couplings of the lightest mass-eigenstate $h_1$ (with mass 120 GeV) 
are expected to correspond to the couplings of  $h$ boson,
whereas couplings of $h_2$ and $h_3$ states can be described
as the superposition of $H$ and $A$ couplings, as follows from table 1.
For the relative basic couplings we have:
\begin{eqnarray}
 \chi^{h_1}_X & \approx  & \chi^h_X  \; ,\nonumber \\
 \chi^{h_2}_X & \approx  & 
        \chi^H_X \cdot \cos \Phi_{HA} \; + \; \chi^A_X \cdot \sin \Phi_{HA}
                                                \; ,     \label{EQ} \\
 \chi^{h_3}_X & \approx  & 
      \chi^A_X \cdot \cos \Phi_{HA} \; - \; \chi^H_X  \cdot \sin \Phi_{HA}
                                                \; ,   \nonumber
\end{eqnarray}
where $X$ denotes a fermion or a vector boson, $X=u,\; d, \; V$.
We study the feasibility of the determination of the mixing angle $\Phi_{HA}$ 
from the combined measurement of the two-photon width and phase for the 
Higgs-boson mass-eigenstate $h_2$.

It should be stressed that, in the considered case of CP violation
via $H-A$ mixing, only the invariant mass distributions are sensitive to
the  mixing angle $\Phi_{HA}$.
In contrast, the angular correlations in Higgs-boson decays  
$higgs \rightarrow WW/ZZ \rightarrow 4f$ can be used to establish
an evidence for direct CP-violation in Higgs-boson couplings \cite{angcor}.

\begin{figure}[p]
  \begin{center}
     \epsfig{figure=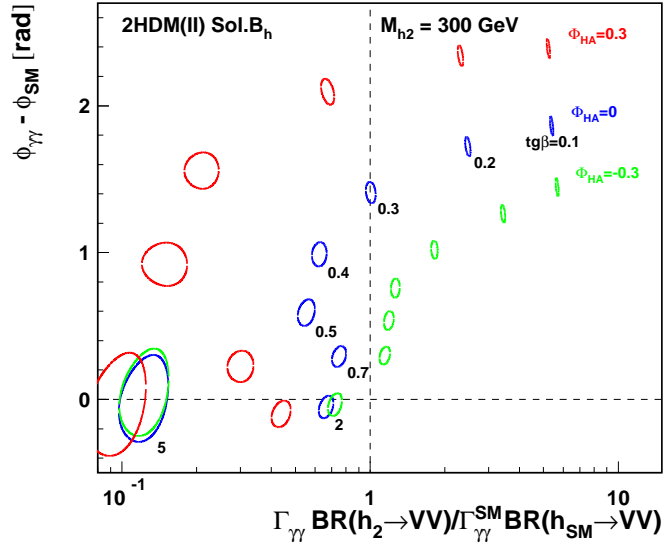,height=\twofigheight,clip=}
  \end{center}
 \caption{  
As in Fig.\ref{fig:showel1} 
for the SM-like 2HDM II (sol. $B_h$) with CP-violation
for    the heavy
 Higgs-boson $h_2$ with mass $ 300$ GeV and couplings 
from Eq.~\ref{EQ}. 
A light Higgs-boson has mass  $M_{h_1} = 120$~GeV.
Three values of $H-A$ mixing angle
$\Phi_{HA}=-0.3,0,0.3$ are considered.
        } 
 \label{fig:showel3} 
 \end{figure}

We perform the combined analysis
of $W^+W^-$ and $ZZ$ decay channels,
of the two-photon width (times vector-boson branching ratio) and 
phase measurement for the scalar Higgs-boson $h_2$ 
with mass $M_{h_2} = 300$ GeV.
Results are presented in Fig.~\ref{fig:showel3}, for
the light Higgs-boson mass  $M_{h_1} = 120$~GeV and $M_{H^{\pm}}=800$ GeV.
Error contours (1$\sigma$) on the measured deviation from
the Standard Model predictions
are shown for    $\Phi_{HA} =0$, i.e. when   CP is conserved,
and for  CP violation with  $\Phi_{HA} = \pm 0.3$~rad.
Even a small CP-violation can significantly influence  the
measured  two-photon width and two-photon phase, and therefore
it is possible to determine precisely  both the CP-violating  mixing 
angle $\Phi_{HA}$ and the parameter $\tan \beta$.

%
%

%
\begin{figure}[p]
  \begin{center}
     \epsfig{figure=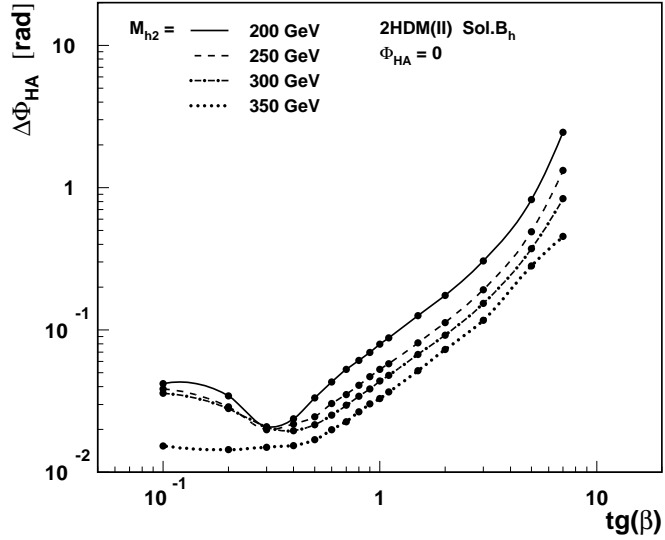,height=\twofigheight,clip=}
  \end{center}
 \caption{ 
    Verifying a   CP-conservation for the 
SM-like 2HDM~II (sol. $B_h$). Statistical error in the determination of 
the $H-A$ mixing angle $\Phi_{HA}$  for four
values of heavy Higgs-boson mass $M_{h_2}$, as a function  of 
 $\tan \beta$ value.
One parameter fit to the observed $W^+ W^-$ and  $ZZ$ mass spectra, 
is considered for the SM-like 2HDM II (sol. $B_h$) with weak CP-violation,
for light Higgs-boson mass of $120$ GeV,
charged Higgs-boson mass of 800~GeV, and $\Phi_{HA}=0$,
Eq.~(\ref{EQ}).
         } 
 \label{fig:errphstat} 
 \end{figure} 
%

Next we address a question: how well can one establish conservation of
CP-symmetry  in the considered model? 
The first estimate can be read out 
from  Fig.~\ref{fig:errphstat} where the statistical error 
in the determination of the angle
$\Phi_{HA}$, around $\Phi_{HA}=0$ value, is shown.
The results are presented  as a function of  $\tan \beta$ 
for four values of Higgs-boson mass $M_{h_2}$, from 200 to 350 GeV.
As above, we assume  a light Higgs-boson mass is equal to 120~GeV
and the charged Higgs-boson mass of 800~GeV.
Here $\Phi_{HA}$ is considered as the only free parameter in the fit. 
Influence of error correlations between $\Phi_{HA}$ and
$\tan \beta$, which have to be taken into account
when both parameters are determined simultaneously 
from the fit, will be discussed in section~\ref{sec:syst}.

The expected statistical error in the determination of $\Phi_{HA}$
is smallest ($\sim$20~mrad) for  $\tan \beta \approx$ 0.3.
For $\tan \beta \sim 1$ the error changes 
from about 30~mrad for mass of 350~GeV to 80~mrad for mass of 200~GeV.
For larger values of $\tan \beta$  the precision of the measurement worsens
fast with increasing $\tan \beta$ value. 
In the considered range of $\tan \beta$ the precision
of phase measurement improves when Higgs-boson mass increases.

%
%

\section{Systematic uncertainties}
\label{sec:syst}

In  sections \ref{sec:2hdm} and \ref{sec:cphiggs}
we considered the statistical errors of the extracted model parameter
($\tan \beta$ or  $\Phi_{HA}$) expected from the one-parameter fit
of the theoretical expectations (cross section convoluted 
with luminosity spectra and detector resolution) 
to the measured invariant-mass distributions 
of $W^+ W^-$ and $\z0 \z0$ events.
However, as a large sample of events is expected, especially
in the $\gamma \gamma \rightarrow W^+ W^-$ channel, systematic
uncertainties can significantly influence the final 
precision and they have to be taken into account. 
In case of 2HDM with CP violation, also the possible correlations
between $\Phi_{HA}$ and $\tan \beta$ have to be considered
if both parameters are used in the fit.

The following sources of the experimental 
systematic uncertainties were considered in the presented analysis:
\begin{itemize}
\item uncertainty in the integrated \gaga\ luminosity
\item uncertainty in the shape of the luminosity spectra
\item uncertainty in the Higgs-boson mass (from other measurements)
\item uncertainty in the total Higgs-boson width
\item energy and mass scale uncertainty of the detector
\item uncertainty in the reconstructed mass resolution
\end{itemize}
In order to take these uncertainties into account we 
allow additional parameters to vary in the fit.
Three model parameters, which were fixed in the previous approach,
are now considered as free parameters:
the integrated \gaga\ luminosity, the Higgs-boson mass 
and the total Higgs-boson width.
To describe uncertainty in the shape of the luminosity spectra,
two new parameters  $A$ and $B$ were introduced,
modifying the CompAZ spectra according to the formula:
\begin{eqnarray*}
 \frac{dL}{dW_{\gamma \gamma}} & = & 
     \frac{dL^{CompAZ}}{dW_{\gamma \gamma}} 
  (1 + A \cdot \sin \pi x + B \cdot \sin 2\pi x) 
\end{eqnarray*}
where $ x = \frac{W_{\gamma \gamma} - W_{min}}{W_{max}-W_{min}} $.
This accounts for possible smooth variations of the luminosity spectra
in the invariant-mass range from $W_{min}$ to $W_{max}$ considered
in the fit.\footnote{The width of the invariant-mass window 
in which the fit was performed changes from 60~GeV
for Higgs-boson mass of 200~GeV to 100~GeV for mass of 350~GeV.}
If the detector energy scale and mass resolution are also considered 
as free parameters in the fit, very large correlations between fitted 
parameters are observed.
This is because, in the limited mass range used for the fit,
deviation of the invariant mass spectra due to the energy scale shift 
is similar to the one resulting from the shift in the Higgs-boson mass.
Similar is true for the mass resolution and the Higgs-boson width.
Therefore, energy scale and mass resolution were fixed in the fit 
as the variations of the Higgs-boson mass and total width 
already account to large extent for possible uncertainties of these parameters.

Variations of the five parameters listed above allow us to account for
possible  deviations of the invariant-mass distributions 
from the nominal model expectation
due to the experimental systematic uncertainties.
We do not impose any additional constraints on these parameters,
which could arise from the independent measurements 
(e.g. luminosity measurement in other process or 
Higgs-boson mass measurement at LC).
Therefore our estimate of systematic effects 
should be considered as the conservative one.

In addition to all experimental uncertainties also 
theoretical uncertainties should be considered.
Unfortunately they are difficult to estimate, 
partly because some of corrections have not yet been calculated. 
Therefore our study is  $K$-factor = 1 type of analysis 
and should be extended by including higher order corrections in the future.

\begin{figure}[p]
  \begin{center}
     \epsfig{figure=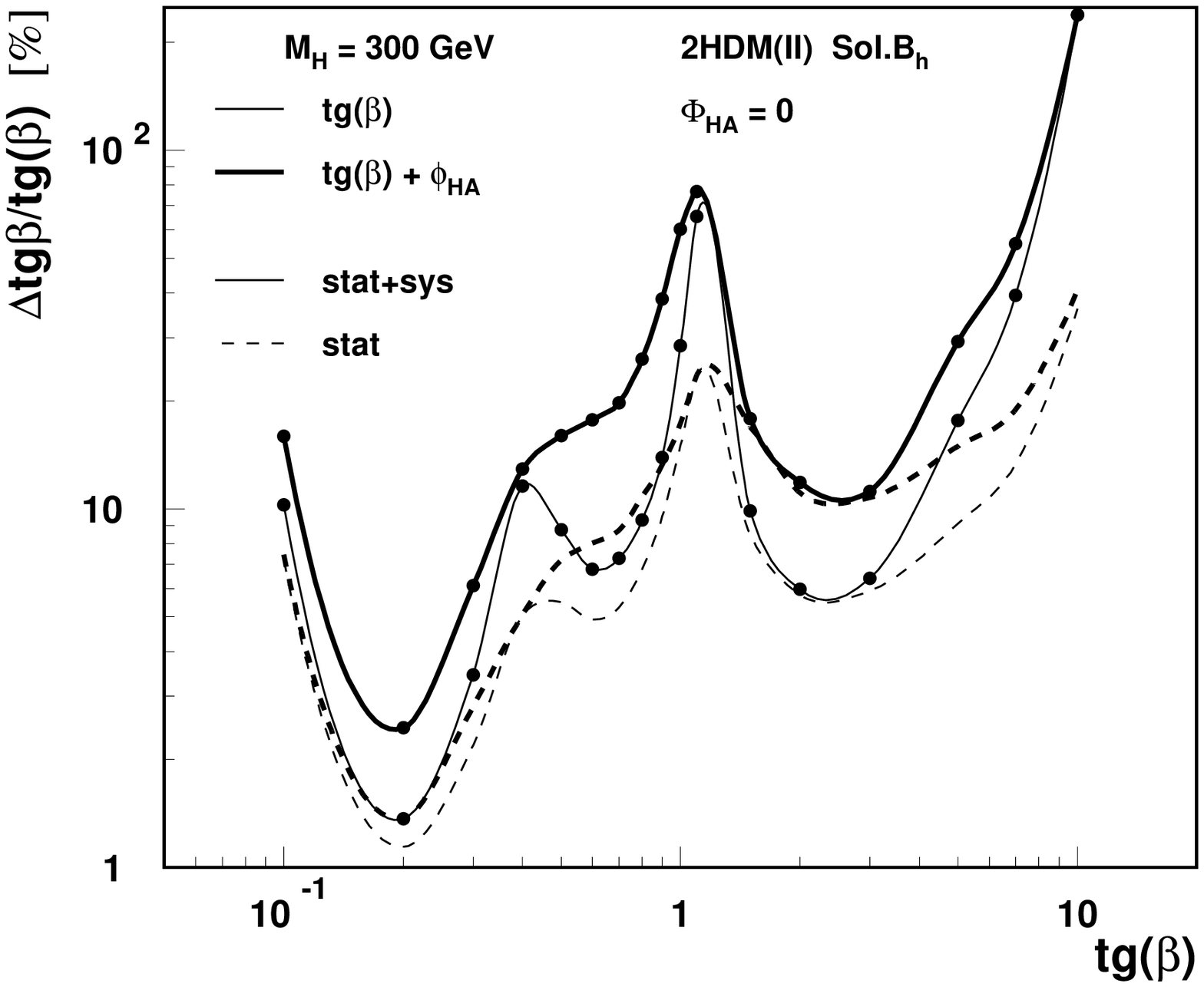,height=\doublefigheight,clip=}
  \end{center}
  \begin{center}
     \epsfig{figure=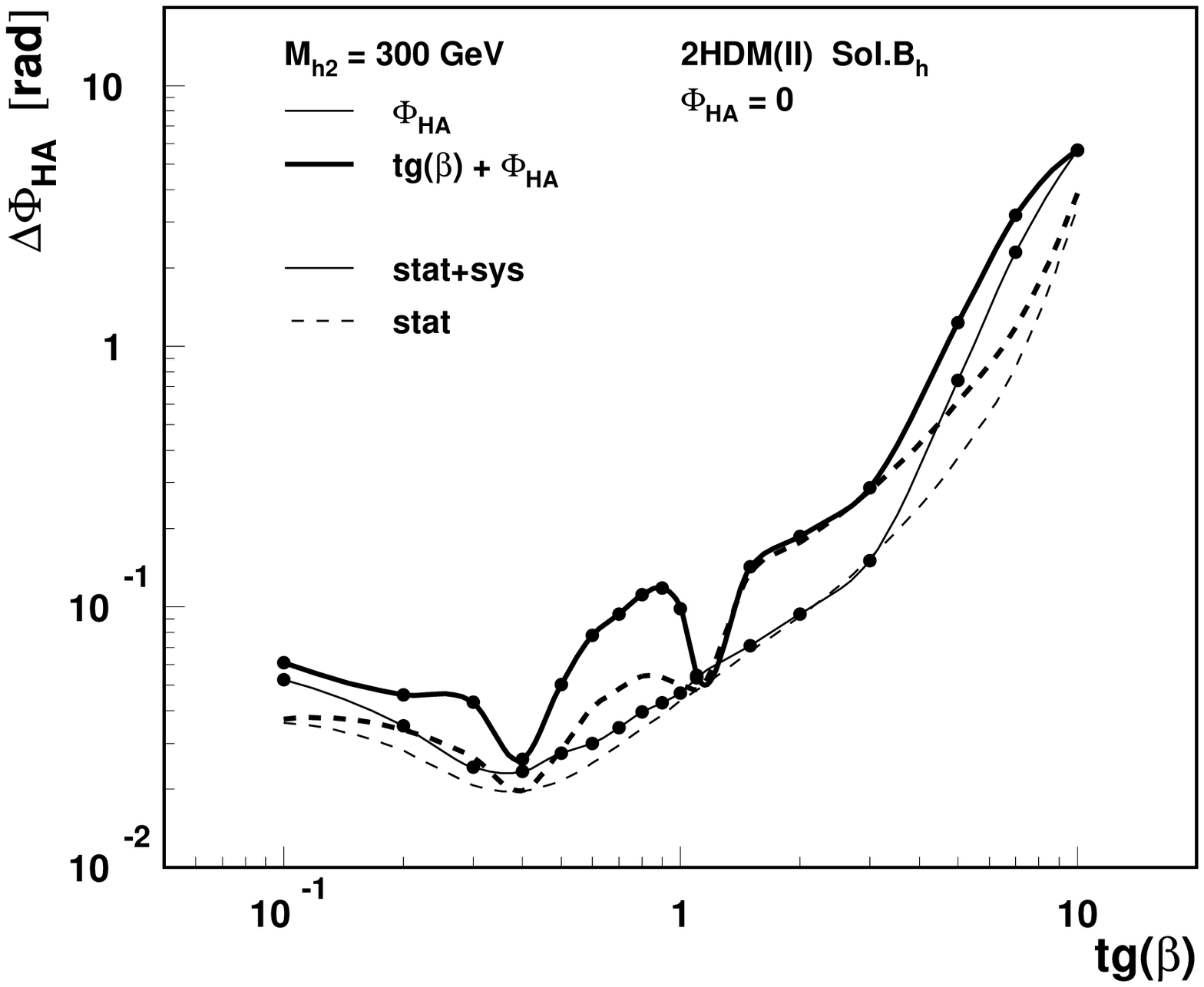,height=\doublefigheight,clip=}
  \end{center}
 \caption{ 
Influence of the systematic uncertainties and parameter correlations
on the expected precision in determination 
of  $\tan \beta$ (upper plot),  
and the mixing angle $\Phi_{HA}$ (lower plot)
for heavy Higgs-boson mass of 300~GeV. 
Errors obtained from the simultaneous fit of both parameters (thick lines)
and from separate fits (thin lines) are compared.
The fit to the observed $W^+ W^-$ and  $ZZ$ mass spectra, 
is considered for the SM-like 2HDM~II (sol. $B_h$), 
with the light Higgs-boson mass of $120$ GeV,
charged Higgs-boson mass of 800~GeV, and no $H-A$ mixing ($\Phi_{HA}=0$),
Eq.~\ref{EQ}.
           } 
 \label{fig:sys2} 
 \end{figure}

In Fig.~\ref{fig:sys2} we present 
the influence of systematic effects on
$\tan \beta$  (upper plot) and $\Phi_{HA}$ (lower plot) measurement,
for a heavy Higgs-boson with mass of 300~GeV.
We assume here the SM-like 2HDM~II (sol. $B_h$), 
with the light Higgs-boson mass of $120$ GeV,
charged Higgs-boson mass of 800~GeV, and no $H-A$ mixing 
(i.e. $\Phi_{HA}=0$).
Errors expected without (dashed lines) and with (solid lines)
systematic uncertainties are compared.
Also shown is the comparison of the errors expected from the simultaneous fit 
of both $\tan \beta$ and $\Phi_{HA}$  (thick lines)
and from separate fits (thin lines).
Systematic uncertainties significantly influence the precision of the
measurement, both for  $\tan \beta$ and $\Phi_{HA}$.
The effect depends strongly on $\tan \beta$ value.
Systematic effects increase the expected error by up to factor of 5
for $\tan \beta$ measurement at the highest $\tan \beta$ values.
Also the correlations between $\tan \beta$ and $\Phi_{HA}$, in the 
simultaneous fit of both parameters, increase the expected errors,
for some cases by factor of 2 or more.
It should also be noted that the effect of the parameter correlations 
is significantly larger when systematic uncertainties are taken into account.

%
%

%
%

\section{Final results including systematic uncertainties}
\label{sec:final}

\begin{figure}[p]
  \begin{center}
     \epsfig{figure=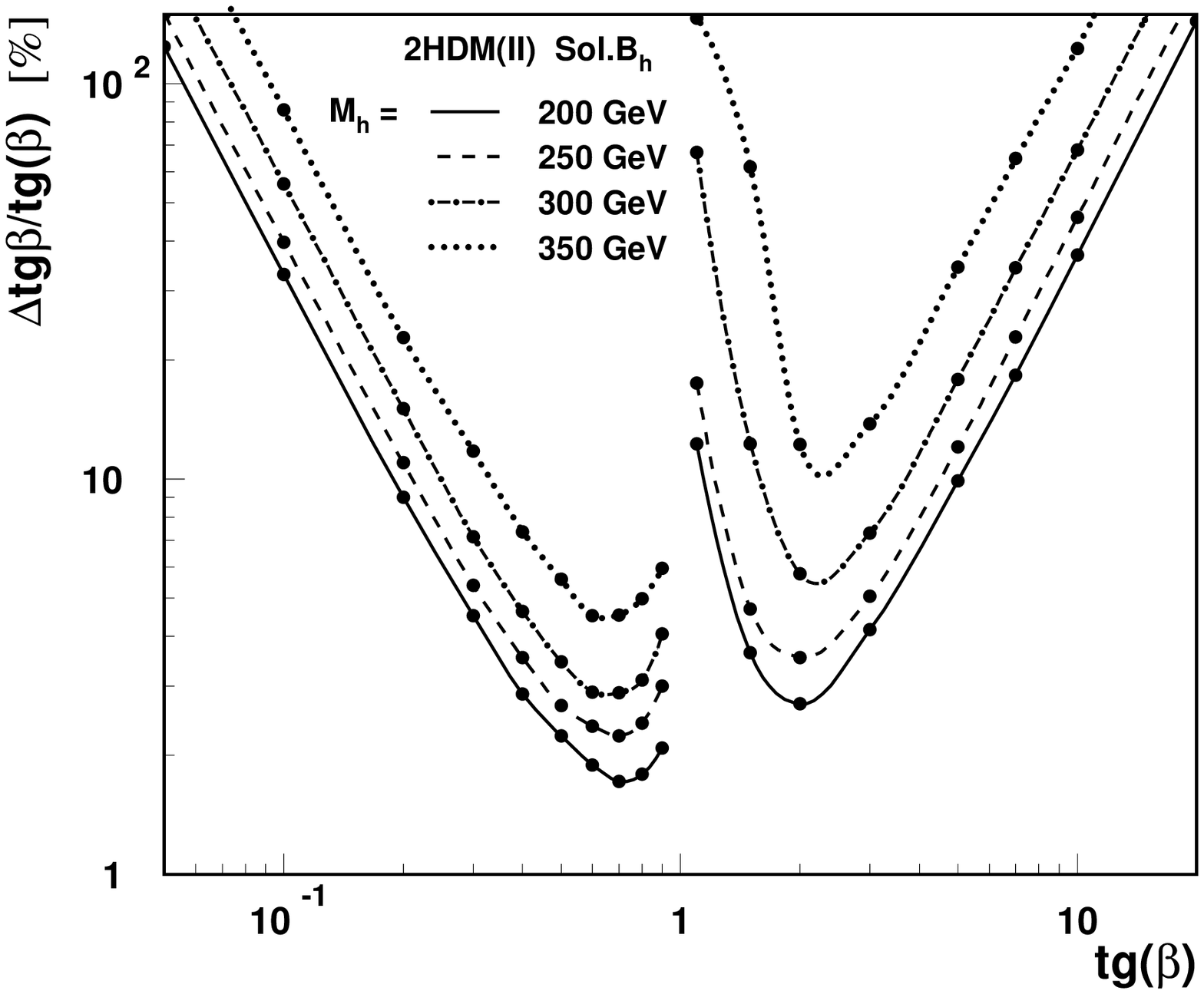,height=\doublefigheight,clip=}
  \end{center}
  \begin{center}
     \epsfig{figure=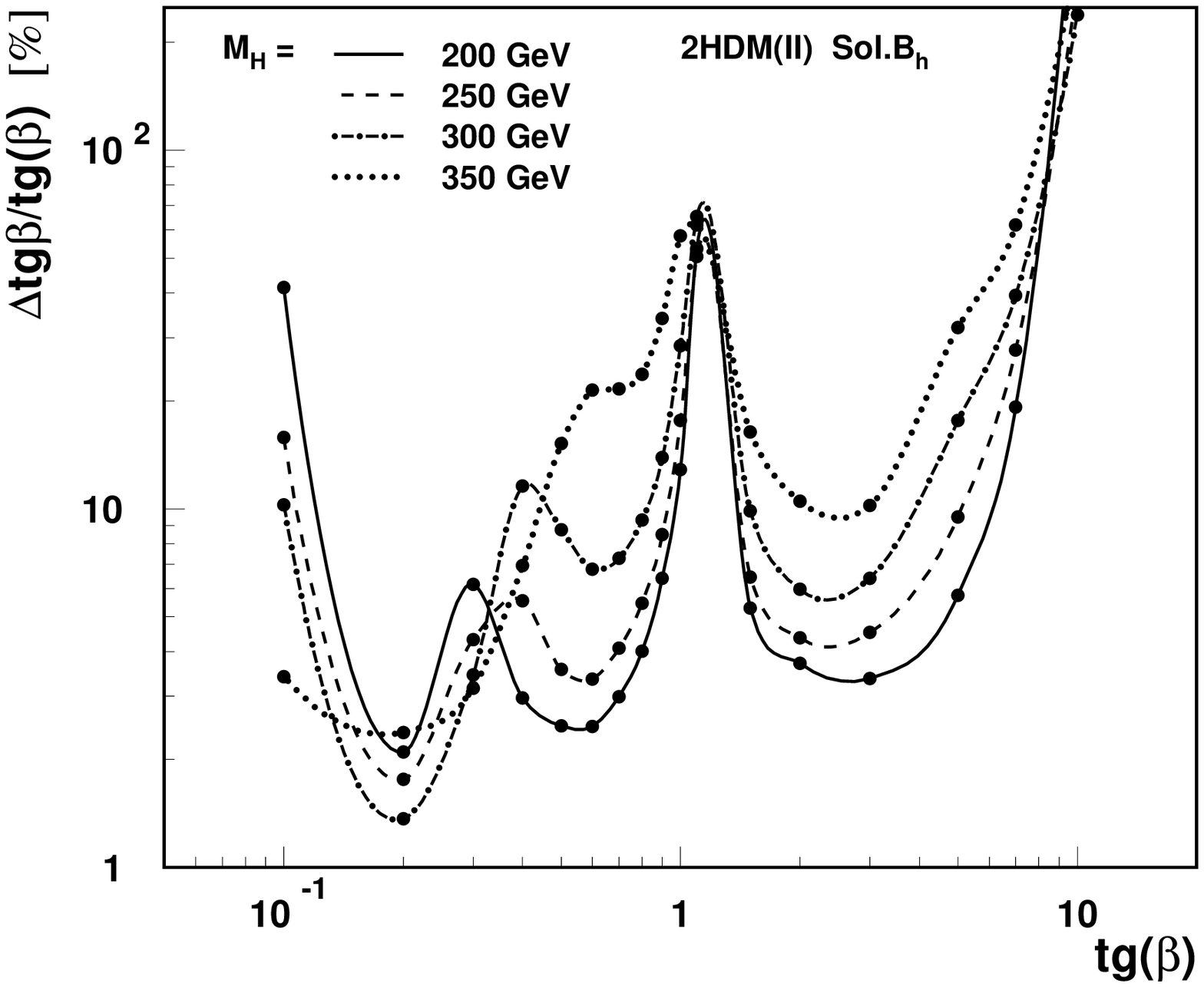,height=\doublefigheight,clip=}
  \end{center}
 \caption{ 
    Total error in the determination of  $\tan \beta$,
   for four values of heavy Higgs-boson mass.
The simultaneous fit to the observed $W^+ W^-$ and  $ZZ$ mass spectra, 
is considered 
for the light Higgs boson (upper plot) and the heavy Higgs boson (lower plot)
of the SM-like 2HDM~II (sol. $B_h$), with charged Higgs-boson mass of 800~GeV.
For measurement of a heavy Higgs boson $H$, 
a light Higgs-boson mass is set to $M_h = 120$ GeV.
Systematic uncertainties related to the luminosity spectra,
Higgs boson mass and total with,
energy scale and mass resolution are taken into account.
           } 
 \label{fig:errsys} 
 \end{figure}

Results presented in  sections \ref{sec:2hdm} and \ref{sec:cphiggs}
have been corrected for the systematic effects and the possible parameter
correlations, as described in section \ref{sec:syst}.
Final results of the analysis, for sol. $B_h$ without CP violation, 
are presented in Fig.~\ref{fig:errsys}.
Total error in the determination of  $\tan \beta$,
as expected from the combined fit to the observed 
$W^+ W^-$ and  $ZZ$ mass spectra, 
is presented for the light (upper plot) and the heavy (lower plot) 
Higgs boson.
In the wide range of the considered   $\tan \beta$ and Higgs-boson mass
values, $\tan \beta$ can be measured with precision better than 10\%.
Although the systematic uncertainties significantly influence the measurement,
the total error of the order of 2\% is still expected 
for most favorable choice of model parameters.

\begin{figure}[p]
  \begin{center}
     \epsfig{figure=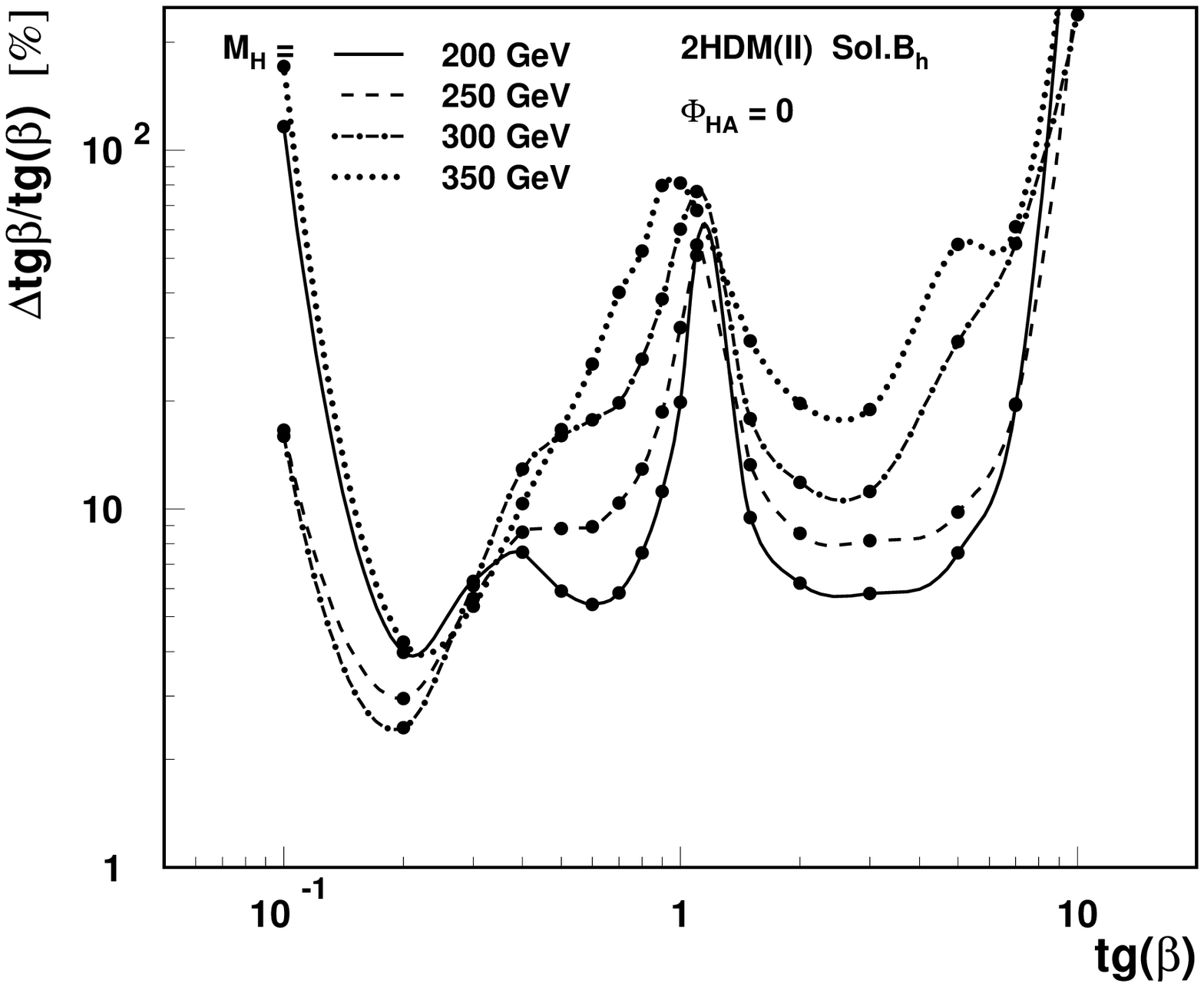,height=\doublefigheight,clip=}
  \end{center}
  \begin{center}
     \epsfig{figure=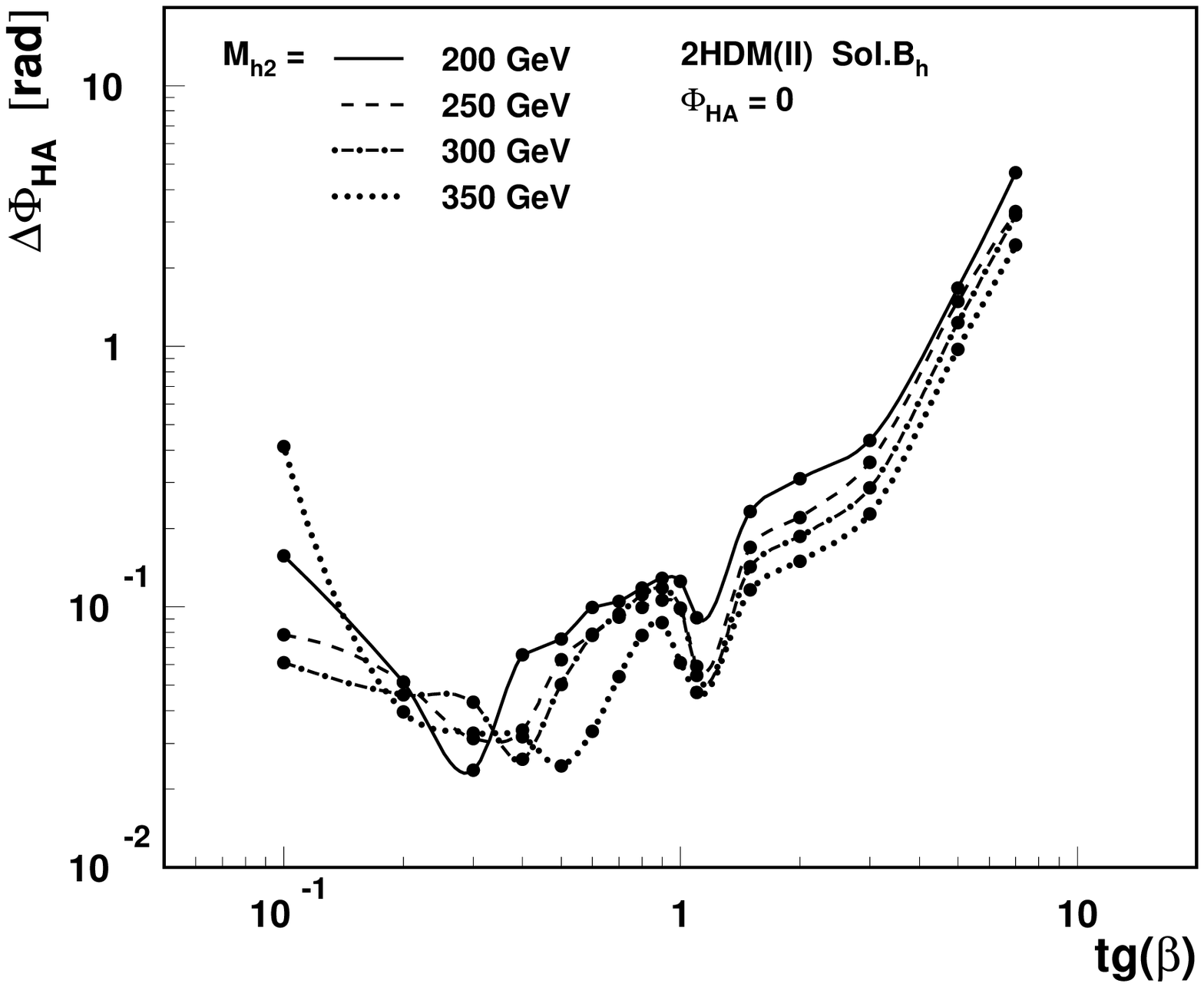,height=\doublefigheight,clip=}
  \end{center}
 \caption{ 
Total error in the determination of 
 $\tan \beta$ (upper plot) and the $H-A$ mixing angle $\Phi_{HA}$ (lower plot),
as a function  of  $\tan \beta$ value, for four
values of heavy Higgs-boson mass $M_{h_2}$. 
The simultaneous fit of both parameters
to the observed $W^+ W^-$ and  $ZZ$ mass spectra, 
is considered for the SM-like 2HDM~II (sol. $B_h$), 
with light Higgs-boson mass of $120$ GeV,
charged Higgs-boson mass of 800~GeV, and no $H-A$ mixing ($\Phi_{HA}=0$),
Eq.~\ref{EQ}.
Systematic uncertainties related to the luminosity spectra,
Higgs boson mass and total with,
energy scale and mass resolution are taken into account.
         } 
 \label{fig:errsys2} 
 \end{figure}

Total errors in the determination of 
$\tan \beta$  and the angle $\Phi_{HA}$,
for sol. $B_h$ with a possible weak CP violation
through a small mixing between $H$ and $A$ states, 
are presented in Fig.~\ref{fig:errsys2},
for four values of heavy Higgs-boson mass.
Errors are evaluated for light Higgs-boson mass of $120$ GeV,
charged Higgs-boson mass of 800~GeV, and $\Phi_{HA}=0$.
The error on $\tan \beta$ increases significantly 
when $\Phi_{HA}$ is included in the fit (compare Fig.~\ref{fig:errsys}).
In most of the considered parameter space it is between 5 and 20\%.
The error on  $\Phi_{HA}$ is below $\sim$100~mrad for $\tan \beta \le 1$
and increases rapidly for high  $\tan \beta$ values.
 
%
%

%
%

We conclude that in the low $\tan \beta$  region 
the assumption that CP-symmetry is conserved in SM-like 2HDM~(II) 
can be precisely verified at the Photon Collider.
However, the fact that the CP-conserving solution $B_h$ fails 
to describe the data,
does not necessarily prove the violation of CP within this model.
Observed discrepancies could also point to the more general 
solution of 2HDM~(II).
In such a case combined analysis of LHC, Linear Collider 
and Photon Collider data is needed to establish a possible
evidence for CP-violation \cite{synergy,lhclc}.

%
%

\section{Conclusions}

The feasibility of measuring the Higgs-boson properties at the 
the Photon Collider at TESLA has been studied in detail for masses 
between 200 and 350~GeV,
using realistic luminosity spectra and detector simulation. 
We consider  the so called solution $B_h$
of the Standard Model-like Two Higgs Doublet Model,  with and 
without CP-conservation.
For the CP conserving case, Yukawa couplings of the lightest Higgs-boson $h$ 
have the same absolute values as in the Standard Model,
and the coupling to the EW gauge-bosons is governed
by only one parameter - $\tan\beta$.
We consider this simple model a generalized SM-like solution $B$ for $h$,
as  the LHC measurement of $h$ production in the gluon-gluon fusion 
process would indicate no deviations from SM.
From the combined measurement of the invariant-mass 
distributions in the $ZZ$ and $W^+ W^-$ decay-channels, the  parameter
of the model can be precisely determined.
After taking into account possible systematic uncertainties
of the measurement, we found out that 
after one year of Photon Collider running
the expected precision in the measurement of the Higgs-boson 
coupling ($\tan \beta$) is of the order of 10\%, 
for both light and heavy scalar Higgs boson.
In case of the Two Higgs Doublet Model solution $B_h$ with a weak CP violation,
the $H-A$ mixing angle can be constrained.
For low $\tan \beta$ values precision 
of about 100~mrad can be obtained (in a small-mixing approximation).


\subsection*{Acknowledgments}
We would like to thank our colleagues from the 
ECFA/DESY study groups for useful comments and suggestions.
This work was partially supported 
by the Polish Committee for Scientific Research, 
grant  no.~1~P03B~040~26
and
project no.~115/E-343/SPB/DESY/P-03/DWM517/2003-2005.
P.N.~acknowledges a partial
support by Polish Committee for Scientific Research, grant 
no.~2~P03B~128~25.
M.K.~acknowledges a partial
support by the European Community's
Human Potential Programme under contract HPRN-CT-2000-00149 Physics
at Colliders.



\end{document}